\documentclass[
    ,final            
  ]
  {aipproc}

\layoutstyle{6x9}

\begin{document}

\centerline{\small{COOL\,05: International Workshop on Beam Cooling, 
Galena, IL \ USA \ 18 -- 23 Sept. 2005}} 
\vspace*{-12pt}

\title{6D Muon Ionization Cooling with an~Inverse~Cyclotron}


\author{D. J. Summers}{
  address={Dept. of Physics and Astronomy, University of Mississippi-Oxford,
University, MS 38677 \ USA}
}

\author{S. B. Bracker}{
  address={Dept. of Physics and Astronomy, University of Mississippi-Oxford,
University, MS 38677 \ USA}
}

\author{L. M. Cremaldi}{
  address={Dept. of Physics and Astronomy, University of Mississippi-Oxford,
University, MS 38677 \ USA}
}

\author{R. Godang}{
  address={Dept. of Physics and Astronomy, University of Mississippi-Oxford,
University, MS 38677 \ USA}
}

\author{R.~B.~Palmer}{
  address={Brookhaven National Laboratory, Upton, NY 11973 \ USA}
}

\begin{abstract}
A large admittance sector cyclotron filled with LiH wedges surrounded by
helium or hydrogen gas is explored. Muons are cooled as they spiral
adiabatically into a central swarm. As momentum approaches zero, the momentum
spread also approaches zero.  Long bunch trains coalesce.  Energy loss is used
to inject the muons into the outer rim of the cyclotron. The density of
material in the cyclotron decreases adiabatically with radius. The sector
cyclotron magnetic fields are transformed into an azimuthally symmetric
magnetic bottle in the center.  Helium gas is used to inhibit muonium
formation by positive muons.  Deuterium gas is used to allow captured negative
muons to escape via the muon catalyzed fusion process. The presence of 
ionized gas in
the center may automatically neutralize space charge. When a bunch train has
coalesced into a central swarm, it is ejected axially with an electric kicker
pulse.
\end{abstract}

\maketitle


\section{Introduction}

Cooling an ensemble of muons
must be completed more rapidly than their 2.2 $\mu$s lifetime.
Ionization cooling can help [1]. Random muon motion is removed by passage
through a low Z material, such as hydrogen, and coherent motion is added with
RF acceleration. Designs for 6D muon cooling using linear
helical channels [2]  at 100 MeV kinetic energies and using frictional
cooling [3,4] at keV energies are under investigation.
Muon cooling rings have been simulated at various levels [5].
In a ring,
the same magnets
and RF cavities may be reused each time a muon orbits. Transverse cooling can
naturally be exchanged for longitudinal cooling by allowing higher
momentum muons to pass through more material. Thus rings cool in all six
dimensions.  

Small emittance bunches of cold muons are useful 
to reduce the aperture of the acceleration system
for a neutrino factory [6,7]
and
are required to provide adequate luminosity for a muon collider [8].
At a neutrino factory, accelerated muons are stored in a
racetrack to produce neutrino beams
($\mu^- \to e^- \, {\overline{\nu}}_e \, \nu_{\mu}$ \, and \,
$\mu^+ \to e^+ \, \nu_e \, {\overline{\nu}}_{\mu}$). Neutrino oscillations
have been observed [9] and need more study. Further
exploration at a neutrino factory could reveal $CP$
violation in the lepton sector [10], and will be  particularly useful if the
$\nu_e$ to $\nu_{\tau}$ coupling, $\theta_{13}$, is small [7,11].
A muon collider can do s-channel scans to split
the $H^0$\,/\,$A^0$ Higgs doublet [12].
Above the ILC's 800 GeV there are a large array of supersymmetric particles
that might be produced [13] and,
if large extra dimensions exist,
so could mini black holes [14].
Note that the energy resolution of a 4 TeV muon
collider is not smeared by beamstrahlung like CLIC.

\section{Operation of an Azimuthally Symmetric Inverse Cyclotron 
at LEAR (p-bar) and PSI (mu-)}

An inverse cyclotron has been used to slow LEAR anti-protons at
CERN [15,16].
An annular quasipotential well, $U(r,z)$,
is formed which ferries anti-protons towards the
center of an azimuthally symmetric cyclotron. The radius of the annulus
decreases with the decreasing angular momentum of the $\overline{p}$.

\begin{equation}
U(r,z) = V(r,z) - (1/(2 \eta r^{\,2}))\, (L_g / M + \eta \,  r A_{\theta})^2,
\end{equation}
where $\eta = e/M$ and $L_g = L_z - e\,r\,A_{\theta}$ is a generalized
angular momentum.  The radial well deepens with decreasing radius and the
vertical well grows shallower (see Fig.~2 of Ref.~15).
Particles must adiabatically spiral to the center.  If dE/dx is too
large, particles will not stay in the magnetic wells.
The final $\overline{p}$ swarm has a radius of 1.5 cm, a height of
4 cm, and a kinetic energy of 2 keV.
A long bunch train is coalesced into a single swarm, which is roughly the
same diameter as the incoming beam.
The spiral time is 20 $\mu$s with
0.3 mbar hydrogen and about 1 $\mu$s with 10 mbar hydrogen.
Given the dependence of the cyclotron frequency on mass,
$f = \omega / 2 \pi = q B / 2 \pi m$,
the spiral time
for a muon is nine time less than for a $\overline{p}$.
The gas pressure in the center must be low, both to allow a particle to
spiral all the way in before
stopping, and to allow reasonable kicker voltages for axial extraction.
An 80 ns electric kicker
pulse rising to 500 V/cm in 20 ns is employed.
The $\overline{p}$'s move 32 cm in 500 ns. Given that
$F = ma$, muons will go nine times farther.

The cyclotron has now been moved from LEAR to PSI where it is
used to slow negative muons to a few keV [17].
Three centimeter diameter beams with 30\,000 $\mu^-$/s below 50 keV
and 0.8 cm diameter beams with 1000 $\mu^-$/s in the 3 to 6 keV kinetic
energy range are output for use.
A static
electric field continuously ejects the muons.  The energy absorber and the
negatively charged 
electrode consist of  a single 30 $\mu$g/cm$^2$ Formvar foil 
(polyvinyl formal) with 3 nm
of nickel produced by 30 minutes of sputtering.

\section{Sketch of a Sector Inverse Cyclotron with Large Admittance for Muons}

A scaling sector cyclotron would allow greater admittance [18] than the
azimuthally symmetric cyclotron now running at PSI. For a given
$\int{\bf{B}}{\cdot}d\,{\bf\ell}$, the ratio of the fields in the hills and
valleys can be adjusted to maximize acceptance.
Only radial and neither spiral nor FFAG [19] sectors have been explored so far.
The sector cyclotron may
be able to function as a damped harmonic oscillator to lower the amplitude of
horizontal and vertical betatron motion as
a bunch train of
muons spirals into a single central swarm.

\begin{figure}[t]
\vspace*{-10mm}
\begin{minipage}{0.58\textwidth}
\centerline{\includegraphics*[width=0.99\textwidth]{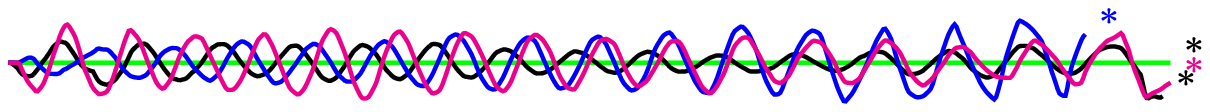}}
\centerline{\includegraphics*[width=0.99\textwidth]{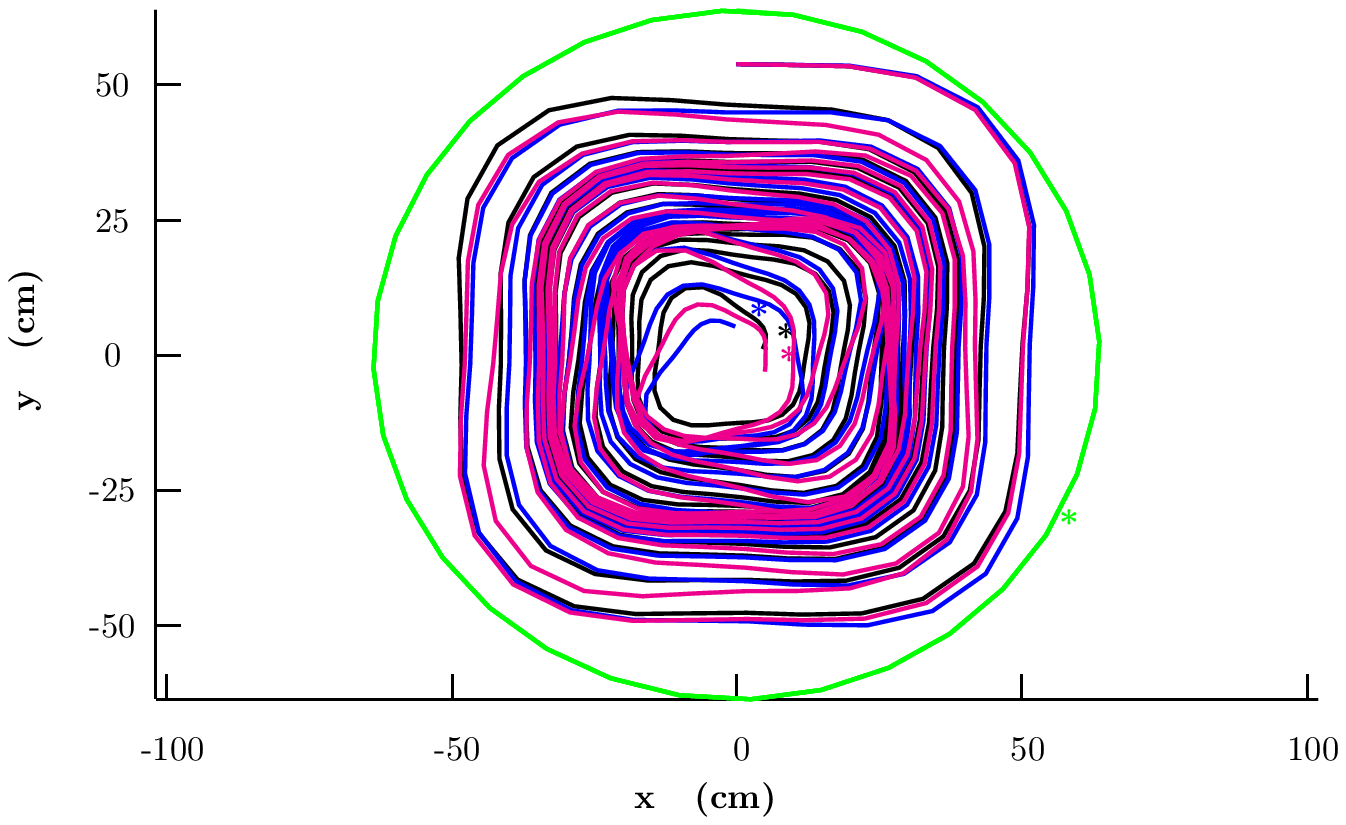}}
\vspace*{-1mm}
\centerline{(a)}
\end{minipage} \hfill
\begin{minipage}{0.39\textwidth}
\centerline{\includegraphics*[width=0.99\textwidth]{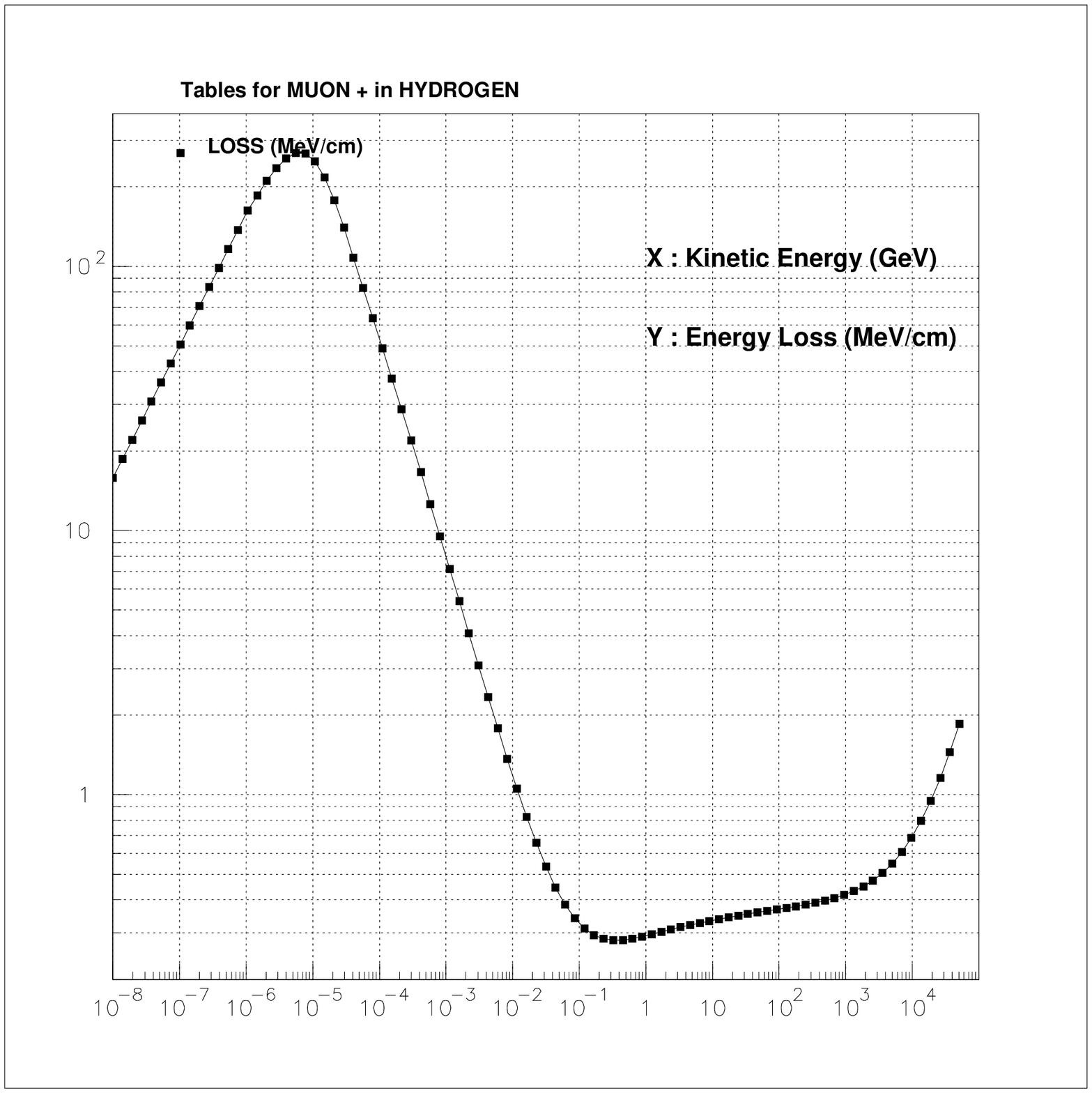}}
\vspace*{-10mm}
\centerline{(b)}
\caption{
{\bf (a)} ICOOL [21] simulation of single turn, energy loss injection.
Three identical 172 MeV/c muons are injected into a 1.8 Tesla cyclotron
with four sectors and soft edged magnetic fields.
The inward spirals differ
because of multiple scattering and straggling. The energy loss is caused by
radial
LiH wedges surrounded by hydrogen gas.  The amount of matter encountered in
a given orbit decreases adiabatically with radius to allow stable orbits.
The upper left trace shows that vertical motion is completely
contained within $\pm$5\,cm along the 70\,m spiral.
The fractional energy loss required in the first turn for injection increases
with the width of the muon beam and decreases as the cyclotron's
magnetic field is lowered.
The injection scaling relation is given by $\Delta{p} = .3 \, B \, \Delta{r}$.
Units are GeV/c, Tesla, and meters, respectively. \break
{\bf (b)} Plot of $\mu^+$ energy loss (MeV/cm) in liquid hydrogen versus kinetic
energy (GeV)
using GEANT3. The default value of ``CUTMUO'' was
decreased
from 10 MeV to 10 eV to propagate slow muons.
The energy turnover at 8 keV corresponds to a momentum of
1.3 MeV/c. $p = \sqrt{2mE} = \sqrt{2 \times 105.7 \times 0.008}$.
Aluminum, copper, iron, and liquid helium show similar results as does the PDG.}
\end{minipage}
\end{figure}

\begin{table}[b]
\caption{The effect of space charge.  Orbital radius of the last muon in 
millimeters is shown as a function of momentum, magnetic field, and central
point charge.  The radii come from Eqn. 2 using
$v = p c^2 / E = p c^2 / \sqrt{p^2 c^2 + m^2 c^4}$.
An ``$i$'' indicates that the radius is partly
imaginary.}
\begin{tabular}{cccccccc} \hline
  & 1 Tesla & 1 Tesla & 1 Tesla \hspace{5mm} & & 2 Tesla & 2 Tesla &2 Tesla \\ 
p & $Q = 0$ & $Q = 10^{12}q$ & $Q = 4 \times 10^{12}q$ & \hspace{5mm} 
  & $Q = 0$ & $Q = 10^{12}q$ & $Q = 4 \times 10^{12}q$\\ \hline
16 MeV/c &$53$  &$27 + 26$    &$27 + 24$     &&$27$  &$13 + 13$  &$13 + 11$\\
8 MeV/c  &$27$  &$13 + 11$     &$13 + 8.7 i$ &&$13$  &$6.7 + 3.6$ &$6.7+9.1 i$\\
4 MeV/c  &$13$  &$6.7 + 9.1 i$ &$6.7 + 22 i$ &&$6.7$ &$3.3 +7.2 i$&$3.3+16 i$\\
2 MeV/c  &$6.7$ &$3.3 + 16 i$ &$3.3 + 32 i$ &&$3.3$ &$1.7 + 11 i$ &$1.7+22 i$\\
1 MeV/c  &$3.3$ &$1.7 + 22 i$ &$1.7 + 45 i$ &&$1.7$ &$.83 + 16 i$ &$.83+32 i$\\
.5 MeV/c &$1.7$ &$.83 + 32 i$ &$.83 + 64 i$ &&$.83$ &$.42 + 23 i$ &$.42+45 i$\\
.25 MeV/c&$.83$ &$.42 + 45 i$ &$.42 + 90 i$ &&$.42$ &$.21 + 32 i$ &$.21+64 i$\\
\hline
\end{tabular}
\end{table}

\begin{equation}
F = {\gamma m\,v^{\,2} \over r} = {qQ \over 4\pi\epsilon_0 r^{\,2}} 
+ q v B, \ \ \
r = {\gamma mv^{\,2} \pm \sqrt{(\gamma m)^{\,2}v^{\,4} 
- 4(qvB)(qQ/4\pi\epsilon_0)} \over 2 q v B}
\end{equation}

\vspace*{-1mm}

With $10^{12}$ muons in a swarm, space charge is a concern.
Table~1 and Eqn.~2 show the effect of space charge. 
Fortunately, the muons are swarming in an ionized gas which may be able to
automatically neuralize the space charge [20]. Electrons experience 200 times
the acceleration of muons in an electric field.
Movement of $10^{12}$ electrons in 100~ns requires 1.6 amps of current.
A metallic grid might also be used for neutralization.

Muons must spiral in fast enough to minimize decay loss, but must not stop
before reaching the central swarm.
So the density of the absorber must decrease smoothly with radius.
Radial LiH wedges immersed in a gas or
high to
low pressure gases separated by beam pipes might meet this criteria. 
The
sector cyclotron geometry must transform into an azimuthally symmetric magnetic
bottle as the muons approach the central swarm.  Otherwise, as shown by GEANT3,
muons will escape though the valleys. In the transition
region the field might resemble a hexapole or octupole field
as used in an
Electron Cyclotron Resonance Ion Source (ECRIS) [22].
If $2 \times 10^{12}$ 172 MeV/c  muons (KE = 96 MeV) arrive at 30\,Hz,
they will deposit 920 watts of beam power.

Atoms can capture muons.  Helium may be used to inhibit muonium 
($\mu^+ e^-$) formation [3].
A possibility for negative muons is to use deuterium gas. Muons will
catalyze fusion and be freed. The sticking factor is 10\%. The reaction appears
in Eqn.~3 [23]. $2 \times 10^{12}$ fusions repeated at 30 Hz only generate
35 watts. The momentum of the freed muon ranges from 0 to 29 MeV/c.
A negatively charged absorber foil might also prevent $\mu^-$ sticking and
is used at PSI. The foil would have to dissipate roughly 100 watts.   

\begin{equation}
d + d + \mu^-\to \, ^3 \! He + n + \mu^- + \, 3.3\,\hbox{MeV} \ \ \hbox{or} \ \ 
t + p + \mu^- + \, 4.0\,\hbox{MeV}
\end{equation}

Busch's theorem (Eqn. 4) [24]
has the effect of increasing the emittance as muons leave a magnetic field.
A half Tesla field and a 50 mm radius give a 4 MeV/c azimuthal kick.
One might be able to use radial iron fins in the exit port to alleviate this
effect or reverse and increase the magnitude of the magnetic field
to capture the unwanted angular momentum in
an absorber after extraction.
Using low fields with tall cylindrical swarms that have small diameters
works for sure.
An RF quadrupole is perhaps a natural choice for acceleration
that would immediately follow the extraction electric kicker.

\begin{equation}
\dot{\phi} = [e / (2 \pi \, \gamma  
\, m \, r^{\,2}(s))] [\Phi(s) - \Phi_k], \quad
L_z = x p_y - y p_x = r^{\,2} \gamma \, m \, \dot{\phi} = -e \, 
B \, r^{\,2} / 2
\end{equation}

\begin{table}[t]
\caption{Emittance reduction goals for an inverse cyclotron.
Emittance goes as $(\Delta p_x \, \Delta x) \, (\Delta
p_y \, \Delta y) \, (\Delta p_z \, \Delta z)$. A muon collider needs a factor
of $10^6$ in cooling.
The input assumes a factor of 10 in transverse cooling [7].
The output for $\Delta{p}$ is from a study of what might be achieved with
frictional muon cooling [3].
}
\begin{tabular}{lcclc} \hline
${\Delta}p_x$  & 30 MeV/c  $\to$ 0.3 MeV/c & \hspace{15mm} &
${\Delta}x$       & 70 mm   $\to$ 50 mm\\
${\Delta}p_y$              & 30 MeV/c  $\to$ 0.3 MeV/c  & &
${\Delta}y$       & 70 mm    $\to$ 50 mm\\
${\Delta}p_z$       & 30 MeV/c $\to$ 0.3 MeV/c & &
${\Delta}z$       & 10000 mm  $\to$ 50 mm   \\
\hline
\end{tabular}
\end{table}
 
In summary, progress on a large admittance sector cyclotron
is underway, including
energy loss injection (see Fig. 1a), 6D muon cooling (see Table 2), 
and an axial electric kicker for extraction.
Many thanks to Juan Gallardo and Franz Kottmann for useful suggestions.
This work was supported by the U.S. Dept. of Energy, DE-FG02-91ER40622
and DE-AC02-98CH10886.

\end{document}